\documentclass[aip,rsi,reprint,twocolumn,superscriptaddress,floatfix]{revtex4-1}

\usepackage{amsmath,amssymb,bm,graphicx}
\usepackage{xcolor}
\usepackage[colorlinks=true,urlcolor=blue,linkcolor=blue,citecolor=blue,bookmarks=false]{hyperref}

\begin{document}

\title{Ultra-high vacuum compatible induction-heated rod casting furnace}

\author{A. Bauer}
\email{andreas.bauer@frm2.tum.de}
\affiliation{Physik-Department, Technische Universit\"{a}t M\"{u}nchen, D-85748 Garching, Germany}

\author{A. Neubauer} 
\affiliation{Physik-Department, Technische Universit\"{a}t M\"{u}nchen, D-85748 Garching, Germany}

\author{W. M\"{u}nzer} 
\affiliation{Physik-Department, Technische Universit\"{a}t M\"{u}nchen, D-85748 Garching, Germany}

\author{A. Regnat} 
\affiliation{Physik-Department, Technische Universit\"{a}t M\"{u}nchen, D-85748 Garching, Germany}

\author{G. Benka} 
\affiliation{Physik-Department, Technische Universit\"{a}t M\"{u}nchen, D-85748 Garching, Germany}

\author{M. Meven} 
\affiliation{Institut f\"{u}r Kristallographie, RWTH Aachen and J\"{u}lich Centre for Neutron Science at MLZ, D-85748 Garching, Germany}

\author{B. Pedersen} 
\affiliation{Heinz Maier-Leibnitz Zentrum (MLZ), D-85748 Garching, Germany}

\author{C. Pfleiderer}  
\affiliation{Physik-Department, Technische Universit\"{a}t M\"{u}nchen, D-85748 Garching, Germany}

\date{\today}

\begin{abstract}
We report the design of a radio-frequency induction-heated rod casting furnace that permits the preparation of polycrystalline ingots of intermetallic compounds under ultra-high vacuum compatible conditions. The central part of the system is a bespoke water-cooled Hukin crucible supporting a casting mold. Depending on the choice of mold, typical rods have a diameter between 6\,mm and 10\,mm and a length up to 90\,mm, suitable for single-crystal growth by means of float-zoning. The setup is all-metal sealed and may be baked out. We find that the resulting ultra-high vacuum represents an important precondition for processing compounds with high vapor pressures under a high-purity argon atmosphere up to 3\,bar. Using the rod casting furnace, we succeeded to prepare large high-quality single crystals of two half-Heusler compounds, namely the itinerant antiferromagnet CuMnSb and the half-metallic ferromagnet NiMnSb. 
\end{abstract}

\maketitle


\section{Introduction}

High-quality single crystals are one of the most important prerequisites for major advances in condensed matter physics. An elegant technique that is widely used for the preparation of single crystals is float-zoning. While the method of heating largely depends on the material to be grown, an essential precondition for float-zoning is the availability of well-shaped and mechanically stable feed rods with a well-defined and homogenous chemical composition. For intermetallic compounds, these rods are typically made by melting the starting elements in an arc furnace or a radio-frequency~(RF) induction-heated furnace. The preparation process, however, may also involve the pressing and sintering of powder components, as common for the growth of oxides.

In order to prepare single crystals of the highest possible purity it is crucial to minimize contaminations in every step of the growth process. In intermetallic compounds, two main sources for contaminants are (i)~the starting elements, where impurity levels of the purest material commercially available typically range from 0.1\,ppm to 1000\,ppm, and (ii)~the high affinity to oxygen, nitrogen, and hydrocarbons of both the starting elements and the final compound. This affinity is drastically enhanced when heating the materials. Therefore, the environment for the preparation of high-quality samples has to be as inert and contaminant-free as possible.

In this paper, we report the design of an all-metal sealed RF induction-heated rod casting furnace~(RCF) that meets these criteria. As its most important facet, the RCF allows to cast polycrystalline rods of metallic elements and intermetallic compounds under ultra-high vacuum~(UHV) compatible conditions. These rods permit the purification of starting elements and serve as feed rods for float-zoning single-crystal growth. In Sec.~\ref{Design}, we present central design aspects of the RCF and describe the typical course of a melting process including the mounting and dismounting of samples. As typical examples, in Sec.~\ref{Materials} we account for the successful preparation of large single crystals of the itinerant antiferromagnet CuMnSb and the half-metallic ferromagnet NiMnSb using the RCF in combination with an UHV-compatible image furnace.\cite{2011:Neubauer:RevSciInstrum}

\section{Design of the rod casting furnace}
\label{Design}

\begin{figure}[t]
\includegraphics[width=1.0\linewidth]{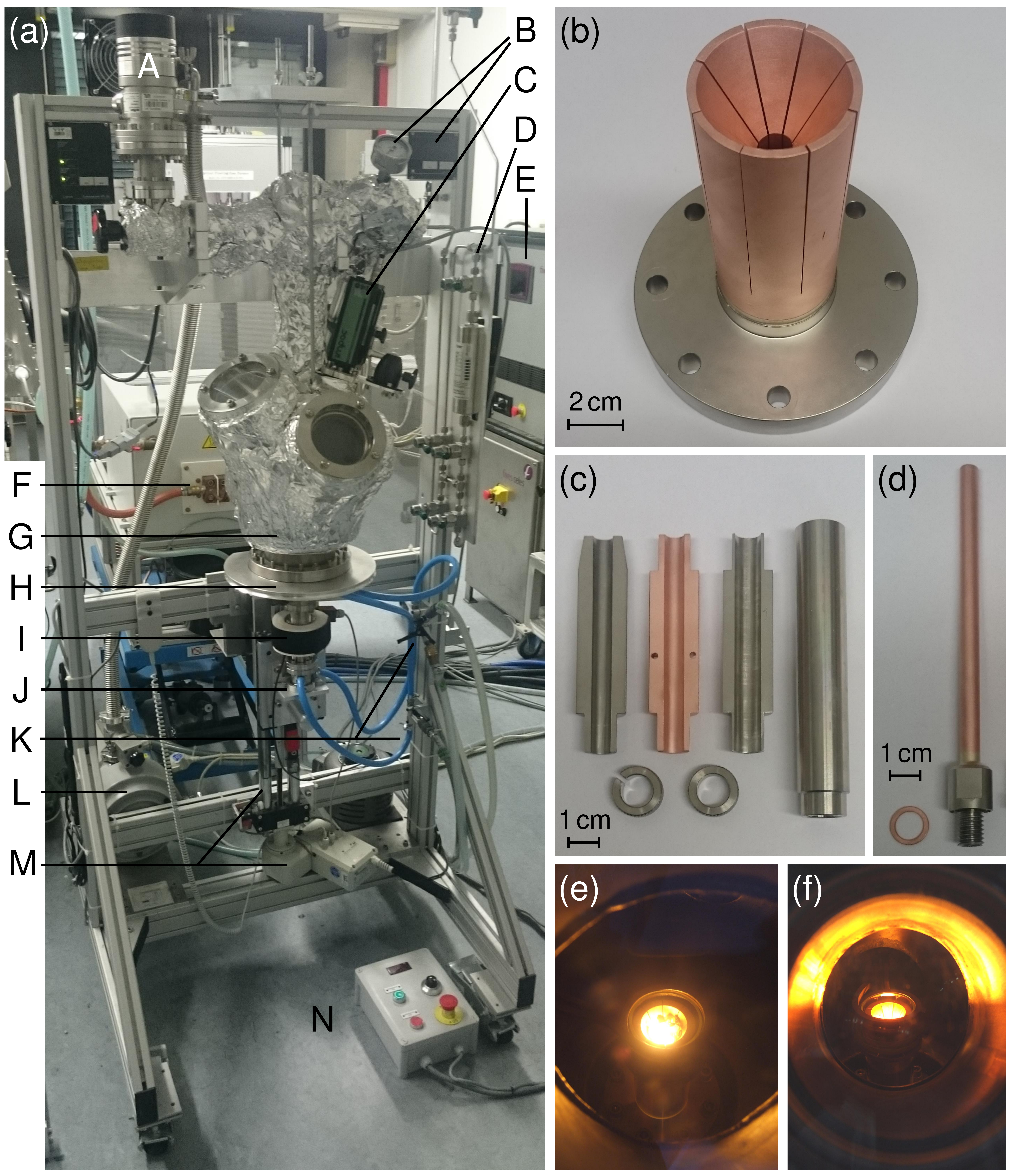}
\caption{UHV-compatible rod casting furnace. (a)~Total view. (A)~Turbomolecular pump, (B)~pressure gauges, (C)~pyrometer, (D)~gas handling and purification system, (E)~RF generator with heat exchanger, (F)~RF capacitor bank, (G)~vacuum chamber with permanently installed heating tapes, (H)~central support flange, (I)~highly flexible metal bellows, (J)~pulling rod on manually moveable sled, (K)~cooling water supply, (L)~roughing pump, (M)~linear actuator for sample mounting, and (N)~RF remote control. (b)~Hukin type copper crucible soldered onto a CF DN63 steel flange. (c)~Different casting molds with clamping rings. (d)~Tip of the pulling rod with copper gasket. (e) and (f)~Left and front viewing port during a melting process.}
\label{figure1}
\end{figure}

\begin{figure}[t]
\includegraphics[width=1.0\linewidth]{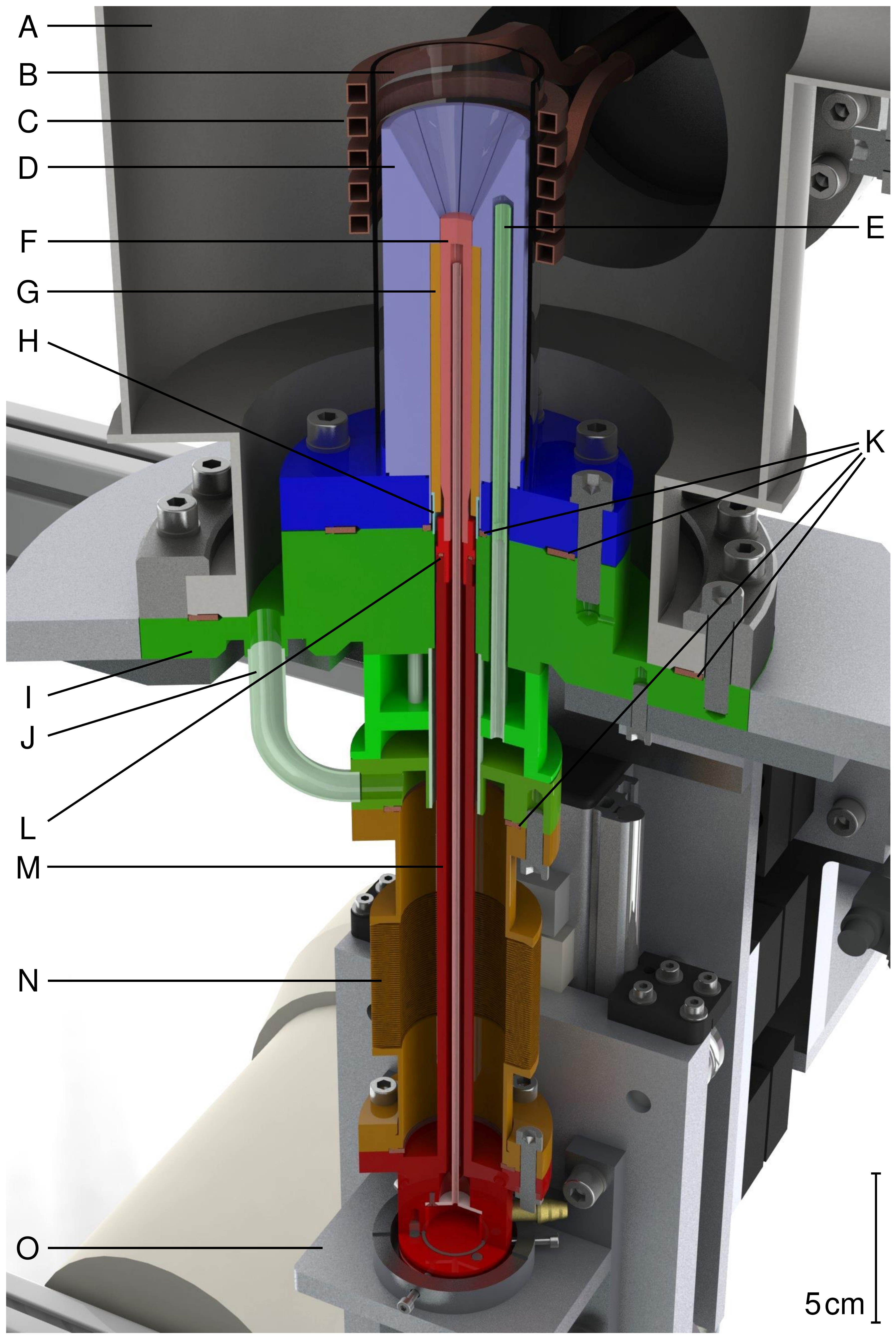}
\caption{Cut-away view of the central section of the rod casting furnace. (A)~Vacuum chamber, (B)~glass tube for electrical insulation, (C)~RF induction coil, (D, blue)~Hukin type copper crucible~(light) soldered onto a steel flange~(dark), (E)~removable steel tubes ensuring desired water flow inside the cooling ducts of the crucible, (F, red)~exchangeable tip of the pulling rod, (G, orange)~casting mold held in place by a (H)~spacer, (I, green)~central support flange providing water cooling and including a (J)~bypass, (K)~standard CF sealings, (L)~bespoke miniaturized copper sealing, (M, red)~bottom part of the pulling rod providing water cooling, (N, brown)~highly flexible metal bellows, and (O)~manually movable sled.}
\label{figure2}
\end{figure}

The RCF consists of a stainless steel vacuum chamber incorporating various bespoke components described in the following. Figs.~\ref{figure1}(a) and \ref{figure2} show a photograph of the entire setup and a cut-away view of its central section, respectively. The core part is a water-cooled Hukin type cold crucible (outer diameter: 45\,mm) as depicted in Fig.~\ref{figure1}(b). The actual copper crucible is high-temperature vacuum-soldered onto a CF DN63 steel flange. On its inside (inner diameter: 16\,mm) the crucible houses a casting mold. Different types of molds may be used, see Fig.~\ref{figure1}(c), where typical examples are one-part or two-part and made of copper or stainless steel. A water-cooled pulling rod completes the sample space of the crucible and prevents the melt from prematurely flowing into the mold. The tip of this pulling rod is exchangeable, see Fig.~\ref{figure1}(d). This feature, in combination with corresponding crucibles and casting molds, permits the configuration of the RCF for the preparation of rods with diameters between 6\,mm and 10\,mm and lengths up to 90\,mm within a few minutes.

The crucible is fastened onto a support flange providing cooling water by eight M8 screws, where two standard CF copper gaskets have to be tightened at the same time in order to seal the water-bearing part. Therefore, the DN16 and DN63 knife edges of the support flange were carefully adjusted to ensure UHV tightness. Inside the cooling ducts of the crucible (diameter: 7\,mm) the desired water flow is achieved through removable steel tubes (outer diameter: 5\,mm). The exchangeable tip of the pulling rod is connected to its base providing cooling water by a bespoke metal seal. This UHV-tight connection is a scaled down version of a CF knife edge and the corresponding copper gasket (outer diameter: 11.5\,mm, height: 1\,mm), tightened by a single M8 thread on the inside of the pulling rod. 

For the preparation of intermetallic compounds high-purity starting elements are placed inside the copper crucible and heated by RF induction. The induction coil possesses a quadratic cross-section, is water-cooled, and is connected to the capacitor bank of a Fives Celes MP 50\,kW generator\cite{FivesCeles} via a CF DN40 RF power feedthrough.\cite{Vacom} A thin-walled quartz glass tube, placed on the steel flange of the crucible, electrically insulates the latter from the RF coil. The electromagnetic levitation arising from the induction reduces the contact between sample and crucible which, together with the small reactivity of the crucible due to its water cooling, minimizes the putative contamination of the sample. Typically, after the starting elements have been molten, the RF heating is turned off. The sample cools down and solidifies inside the crucible. The resulting pill is flipped by pushing the pulling rod upwards and is subsequently remelted. Repeating this process several times ensures excellent compositional homogeneity of the intermetallic ingots. As the last step, the pill is melted one final time before a sudden downward movement of the pulling rod allows the melt to flow into the casting mold forming a polycrystalline rod. The required vertical movement of the pulling rod of about 100\,mm is made possible by a highly flexible COMVAT metal bellows\cite{Comvat} in combination with a sled guiding the motion. 

In order to monitor the process from various angles, the design of the vacuum chamber includes three CF flanges with Spectrosil 2000 high-pressure quartz glass windows. The view through the DN100 windows on the left and on the front of the chamber during the heating of a sample is depicted in Figs.~\ref{figure1}(e) and \ref{figure1}(f), respectively. The DN40 viewing port on the top of the chamber permits to measure the temperature of the sample using a pyrometer, namely, an IMPAC IGA 140 MB 25 L\cite{LumaSense} with a temperature range from 350\,$^{\circ}$C to 2500\,$^{\circ}$C. When precise temperature values are required, the measurement may by corrected for the emissivity of a given material and the absorption by the glass. In most cases, however, the reading of the pyrometer only serves a rough indicator for thetemperature of the sample.

To dismount the cast rod and mount new starting elements, the DN160 seal connecting the support flange and the vacuum chamber is opened. Subsequently, the lower part of the furnace, including the crucible and its water support as well as the pulling rod and the sled, is lowered by about 300\,mm with a Rose+Krieger EPX30 linear actuator.\cite{RoseKrieger} After emptying the cooling water circuit using pressurized air, the crucible may be removed. Unscrewing the eight cooling water tubes, cf.\ Fig.~\ref{figure2}(E), provides easy access to the casting mold and the tip of the pulling rod. By fixing the sled in its upmost position, the tip of the pulling rod may be exchanged without detaching its base from the system. Following a thorough cleaning of all parts, which usually involves glass bead blasting, chemical etching, and ultrasonic baths, the system may be reassembled.

In order to reach UHV prior to a high-purity sample preparation, the furnace is evacuated with a Leybold Turbovac 50 turbomolecular pump backed by an oil-free Varian SH-110 scroll pump. The all-metal sealing allows to bake the system using permanently installed heating tapes on the vacuum chamber and a bespoke heating jacket for the metal bellows. This way final pressures in the range of 10$^{-8}$\,mbar are reached within 24\,h. When handling compounds with high vapor pressure, UHV represents a precondition for the use of a high-purity argon atmosphere equivalent to UHV conditions. For this purpose, 6N argon gas of up to 3\,bar is additionally purified with a point-of-use gas purifier\cite{SAESgetter} leading to nominal impurity levels below 1\,ppb.

\section{Materials preparation}
\label{Materials}

\begin{figure} 
\includegraphics[width=1.0\linewidth]{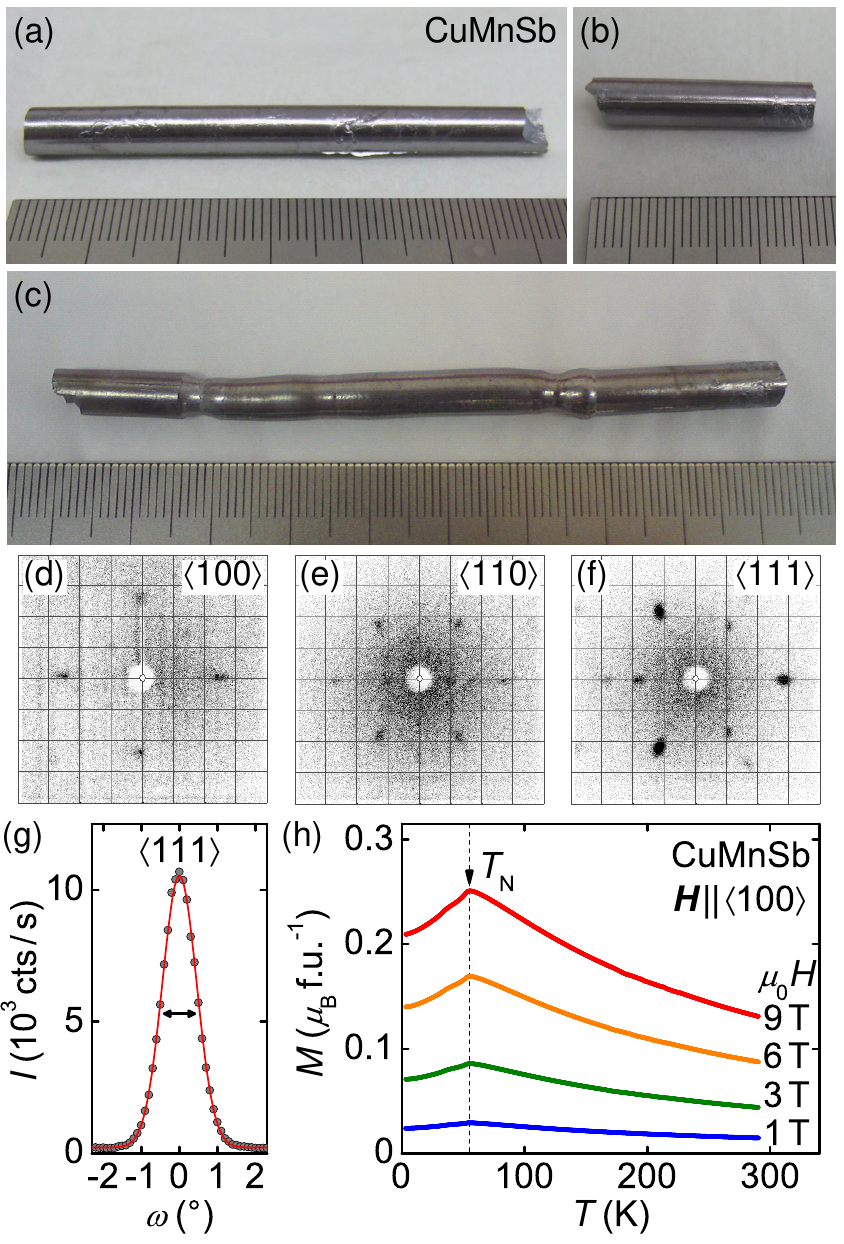}
\caption{Single-crystal growth of CuMnSb. (a)~Polycrystalline feed rod cast with the RCF. (b)~Polycrystalline seed rod. (c)~Optically float-zoned single crystal. Growth direction was from right to left. \mbox{(d)--(f)}~Laue X-ray backscattering pictures from polished $\langle100\rangle$, $\langle110\rangle$, and $\langle 111\rangle$ surfaces, respectively. (g)~Single-crystal neutron diffraction data around a crystalline $\langle 111\rangle$ reflection at $T = 65$\,K. Error bars are smaller than the symbol size. (h)~Magnetization as a function of temperature for various magnetic fields applied along a crystallographic $\langle100\rangle$ axis.}
\label{figure3}
\end{figure}

\begin{figure} 
\includegraphics[width=1.0\linewidth]{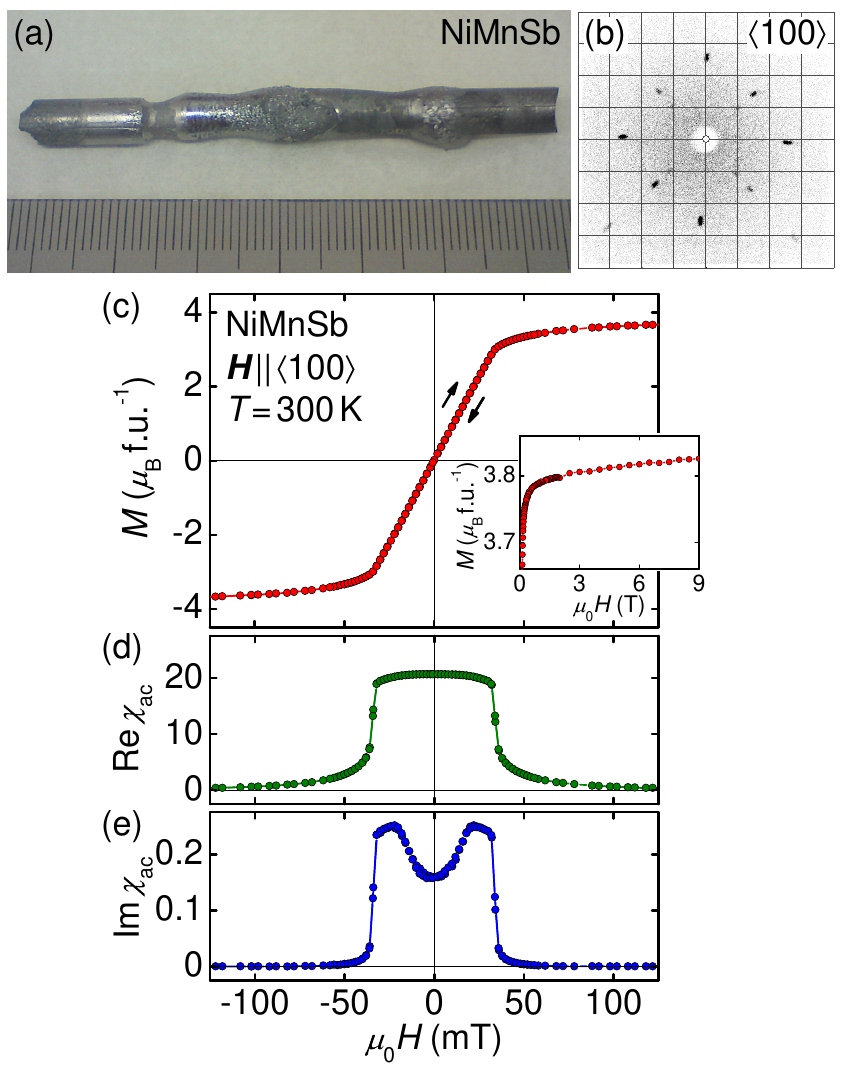}
\caption{Single-crystal growth of NiMnSb. (a)~Optically float-zoned single crystal. Growth direction was from right to left. (b)~Laue X-ray backscattering picture from a polished $\langle100\rangle$ surface. (c)~Magnetization as a function of applied magnetic field along a crystallographic $\langle 100\rangle$ axis at room temperature. Note the absence of hysteresis on the scale of this study. The inset shows that the magnetization stays unsaturated up to 9\,T, the highest field value studied. (d) and (e)~Magnetic field dependence of the real and the imaginary part of the ac susceptibility.}
\label{figure4}
\end{figure}

The RCF is mostly used to cast polycrystalline feed rods of intermetallic compounds. By now more than two hundred rods were synthesized and subsequently float-zoned to single crystals. The range of materials includes for instance Heusler and half-Heusler compounds,\cite{2010:Neubauer:PhD, 2011:Neubauer:RevSciInstrum, 2012:Neubauer:NuclInstrumMethodsPhysResA, 2015:Hugenschmidt:ApplPhysA, 2015:Weber:PhysRevLett} cubic chiral magnets with B20 structure,\cite{2008:Munzer:Diploma, 2010:Pfleiderer:JPhysCondensMatter, 2010:Munzer:PhysRevB, 2010:Bauer:PhysRevB, 2012:Bauer:PhysRevB} as well as various binary and ternary rare earth compounds. We further note that for both the purification of starting elements and the preparation of high-quality intermetallic feed rods, it is advantageous that oxides in general possess higher melting temperatures and lower mass densities compared to the corresponding pure metals. In cases where the starting elements are heavily contaminated, we find that oxides tend to float as solid flakes on top of the molten metallic sample. After casting a rod from the melt, these oxide flakes remain as a slag in the upper part of the crucible.

In the following, we illustrate the potential of the RCF in the context of the preparation of two half-Heusler compounds. The first example is CuMnSb, a low-temperature antiferromagnet\cite{1952:Nowotny:MonatshChem, 1952:Castelliz:MonatshChem, 1968:Endo:JPhysSocJpn, 1968:Forster:JPhysChemSolids} that shows characteristics of both itinerant and local-moment magnetism.\cite{2003:Pfleiderer:PhysicaB, 2004:Doerr:PhysicaB, 2006:Boeuf:PhysRevB} Recent band structure calculations identify CuMnSb as a compensated semi-metallic compound in which the antiferromagnetic phase can be pictured heuristically as self-doped Cu$^{1+}$Mn$^{2+}$Sb$^{3-}$.\cite{2005:Jeong:PhysRevB} It is the only antiferromagnet among $3d$-based half-Heusler compounds and the comparably low transition temperature indicates the importance of spin fluctuations, making CuMnSb a fascinating candidate for further studies. The second example is the half-metallic high-temperature ferromagnet NiMnSb.\cite{1951:Castelliz:MonatshChem} In this compound full spin polarization at the Fermi surface is a broadly accepted property of the bulk material, where the states at the Fermi level exhibit strong Mn character.\cite{1983:deGroot:PhysRevLett, 1986:Hanssen:PhysRevB, 1990:Hanssen:PhysRevB, 2011:Graf:ProgSolidStateCh}

The preparation of the single crystals was similar for both materials. First, commercially available manganese flakes~(4N), typically covered by a thick oxide layer, were etched using nitric acid, thoroughly cleaned, and cast into a solid rod under argon atmosphere in the RCF. While etched pieces of manganese tarnish within minutes, potentially triggered by remaining contaminations or their large surface area, we find empirically that the surface of cast rods stays metallically clean even under ambient conditions. Second, the purified manganese was combined with antimony pieces~(6N) and copper shot~(5N) or nickel pellets~(4N5), respectively. Subsequently, as depicted in Fig.~\ref{figure3}(a), a polycrystalline feed rod was cast under argon atmosphere in the RCF as described above. A second rod, see Fig.~\ref{figure3}(b), was obtained by repeating the procedure. Third, these two rods were float-zoned into a single crystal in our UHV-compatible image furnace.\cite{2011:Neubauer:RevSciInstrum} The resulting ingots of CuMnSb and NiMnSb are shown in Figs.~\ref{figure3}(c) and \ref{figure4}(a), respectively. Laue X-ray backscattering pictures as well as polished cuts throughout the float-zoned sections confirmed their single crystallinity after an initial grain selection. Note that copper fluorescence prohibits better quality Laue pictures for CuMnSb, cf.\ Figs.~\ref{figure3}(d) through \ref{figure3}(f). Single-crystal neutron diffraction, however, performed on the diffractometers RESI\cite{2015:Pedersen:JLSFR} and HEiDi\cite{2015:Meven:JLSFR} at the Heinz Maier-Leibnitz Zentrum~(MLZ) confirmed the high crystalline quality. As an example, we show the profile of a crystalline $\langle 111\rangle$ reflection in Fig.~\ref{figure3}(g) as measured on HEiDi at an incident neutron wavelength of 0.79\,\AA. We determine a full width at half maximum of $1^{\circ}$ indicating a small mosaicity for this type of compounds.

The magnetization of CuMnSb was measured in an Oxford Instruments vibrating sample magnetometer. The magnetization of NiMnSb was measured in a Quantum Design physical properties measurement system by means of a standard extraction method. In addition, ac susceptibility was determined at an excitation frequency of 911\,Hz and an excitation amplitude of 1\,mT. For both CuMnSb and NiMnSb, single-crystalline bars of $6\times1\times1\,\mathrm{mm}^{3}$ were used with their long edge and the applied magnetic field parallel to an $\langle100\rangle$ axis. The small edges of the samples were parallel to $\langle110\rangle$ axes. 

Fig.~\ref{figure3}(h) shows the temperature dependence of the magnetization of CuMnSb for applied fields of 1\,T, 3\,T, 6\,T, and 9\,T. At a N$\acute{\textrm{e}}$el temperature of $T_{\mathrm{N}} = 55$\,K the magnetization exhibits a clear kink that does not shift in the field range studied. This observation is consistent with measurements on polycrystals reported before.\cite{1970:Endo:JPhysSocJpn, 1984:Helmholdt:JMagnMagnMater, 2006:Boeuf:PhysRevB} From a Curie-Weiss fit we extract a fluctuating moment of $4.0\,\mu_{\mathrm{B}}\,\mathrm{f.u.}^{-1}$. The magnetic properties are essentially isotropic where the magnetization for field parallel to $\langle100\rangle$, $\langle110\rangle$, and $\langle111\rangle$ differs by less than 5\% (not shown).

Polycrystals of CuMnSb are prone to the formation of the impurity phase Mn$_{2}$Sb\cite{1970:Endo:JPhysSocJpn, 2006:Boeuf:PhysRevB} which is ferrimagnetic\cite{1957:Wilkinson:JPhysChemSolids} and leads to a pronounced maximum in susceptibility and magnetization around 200\,K. The lack of the latter in our data demonstrates the absence of this impurity phase in our samples. Further details on the metallurgical characterization as well as on the magnetic and transport properties of CuMnSb are reported elsewhere.\cite{2015:Bauer:PhD, 2016:Regnat:PhD} Our samples also allow for detailed single-crystal elastic as well as inelastic neutron scattering studies.\cite{2016:Brandl:PhD}

Fig.~\ref{figure4}(c) depicts the room temperature magnetization of NiMnSb as a function of applied magnetic field along the crystalline $\langle100\rangle$ axis. We observe no hysteresis on the field scale studied. The ordered moment at 300\,K is $3.8\,\mu_{\mathrm{B}}\,\mathrm{f.u.}^{-1}$, in perfect agreement with literature\cite{1996:Hordequin:JMagnMagnMater}. Note that the moment stays unsaturated up to 9\,T, the highest field value studied.

\section{Conclusions}

In conclusion, we reported the design of an all-metal sealed and RF induction-heated rod casting furnace that permits the preparation of polycrystalline rods of intermetallic compounds under UHV-compatible conditions. Since the furnace was put into operation, more than two hundred rods of various materials were cast. Most of them, as exemplary illustrated on the half-Heusler compounds CuMnSb and NiMnSb, were subsequently float-zoned to large single crystals.

\acknowledgments

We wish to thank M.~Pfaller and the team of the Zentralwerkstatt of the Physik-Department at TUM. We gratefully acknowledge discussions and support by P.~B\"{o}ni, R.~Bozhanova, G.~Benka, H.~Ceeh, A.~Erb, S.~Giemsa, M.~Halder, D.~Mallinger, S.~Mayr, J.~Remmel, S.~Gottlieb-Sch\"{o}nmeyer, B.~Russ, and R.~St\"{o}pler. We are especially indebted to G\"{u}nter Behr, IFW Dresden. Neutron single-crystal diffraction data were provided by RESI operated by LMU/TUM and by HEiDi operated by RWTH Aachen and JCNS (JARA collaboration) at MLZ. Financial support through the Deutsche Forschungsgemeinschaft (DFG) and ERC Advanced Grant 291079 (TOPFIT) are gratefully acknowledged. A.B., A.R., and G.B.\ acknowledge financial support through the TUM graduate school.

\bibliography{RCF}

\end{document}